\documentclass[aps,pra,reprint,superscriptaddress,showpacs]{revtex4-1}
\usepackage[dvipdfmx]{graphicx}
\usepackage{subfigure}
\begin{document}

\title{Spin Hall Effect in a Spinor Dipolar Bose-Einstein Condensate}
\author{T. Oshima}
\affiliation{Department of Applied Physics, University of Tokyo, Tokyo 113-8656, Japan}
\author{Y. Kawaguchi}
\affiliation{Department of Applied Physics, Nagoya University, Nagoya 464-8603, Japan}
\date{\today}
\begin{abstract}
We theoretically show that the spin Hall effect arises in a Bose-Einstein condensate (BEC) of neutral atoms interacting via the magnetic dipole-dipole interactions (MDDIs). Since the MDDI couples the total spin angular momentum and the relative orbital angular momentum of two colliding atoms, it works as a spin-orbit coupling. Thus, when we prepare a BEC in a magnetic sublevel $m=0$, thermally and quantum-mechanically excited atoms in the $m=1$ and $-1$ states feel the Lorentz-like foces in the opposite directions.  This is the origin for the emergence of the the spin Hall effect. We define the mass-current and spin-current operators from the equations of continuity and calculate the spin Hall conductivity from the off-diagonal current-current correlation function within the Bogoliubov approximation. We find that the correction of the current operators due to the MDDI significantly contributes to the spin Hall conductivity. Possible experimental situation is also discussed.
\end{abstract}
\maketitle
\section{Introduction}
Spin-orbit interaction (SOI) is one of the key ingredients for the emergence of non-trivial transport phenomena such as the anomalous Hall effect\cite{RevModPhys.82.1539}, the spin Hall effect\cite{science301,PhysRevLett.92.126603}, and robust surface/edge states of topological insulators\cite{RevModPhys.83.1057}. The spin-orbit coupling phenomena are also intensively investigated using cold atomic systems\cite{SOICIQG,RevModPhys.83.1523}, since the synthetic spin-dependent magnetic field has been experimentally realized\cite{PhysRevLett.102.130401,electric,Ylin1,Ylin3,PhysRevLett.107.255301}. For example, various spin textures are predicted to appear in a harmonically trapped Bose-Einstein condensate (BEC) with the one- to three-dimensional SOIs\cite{PhysRevLett.105.160403,PhysRevA.83.043616,PhysRevA.83.053602,PhysRevA.84.011607,PhysRevLett.107.200401,PhysRevLett.108.010402,PhysRevLett.107.270401,PhysRevLett.109.015301,PhysRevA.88.013612,PhysRevA.91.043616,PhysRevA.91.053624}, where the experimental realizations of the spin-orbit coupled spin-1 system\cite{2015arXiv150105984C} and the two-dimensional SOI\cite{2015arXiv151108170W} have recently been reported. The coupling between spin and orbital degrees of freedom directly results in the observation of the spin Hall effect\cite{spin}. When atoms are confined in an optical lattice, the synthetic gauge field is expected to cause topological band structures as in the case of solid-state materials. By utilizing the unprecedented tunability of cold atomic systems, the topological properties of the band structure have been experimentally investigated\cite{DMOZH,OCC,ERTHM,MCNHB,ABI,Mancini1510}.

Although the synthetic SOI in cold atomic systems induces such interesting phenomena, they are basically understood from the single-particle physics, because the synthetic SOI is the one-body interaction of atoms with a static (non-Abelian) gauge field. Contrarily to this, the magnetic dipole-dipole interaction (MDDI) is regarded as a two-body SOI: The MDDI is the inter-atomic interaction between atoms with magnetic moments and couples the total spin angular momentum and the relative orbital angular momentum of two colliding atoms. In general, the MDDI becomes prominent in atomic gases with large magnetic moments, such as Cr\cite{PhysRevLett.94.160401}, Dy\cite{PhysRevLett.107.190401}, and Er\cite{PhysRevLett.108.210401}, where the long-range and anisotorpic nature of the MDDI causes exotic phenomena\cite{baranov2012condensed,0034-4885-72-12-126401}, some of which were experimentally observed\cite{SDE,PhysRevLett.101.080401,ORI,2016arXiv160103318F,PhysRevLett.109.155302,PhysRevLett.111.185305,PhysRevLett.108.215301,Aikawa1484}. Even for alkali-metal atoms, by fine tuning the experimental parameters the MDDI-induced phenomena, such as the decoherence of an atomic interferometer\cite{PhysRevLett.101.190405}, the deformation of the condensate\cite{PhysRevLett.102.090402}, inhomogeneous Larmor precession due to the dipolar filed\cite{PhysRevLett.112.185301}, and the emergence of the magnon energy gap\cite{PhysRevLett.113.155302}, have been observed. Since the MDDI couples the spin and orbital degrees of freedom, it is also known to contribute to the magnetization relaxation\cite{Hensler2003,DCG,PhysRevLett.106.255303,PhysRevLett.108.045307}. In particular, when we prepare a BEC in ultralow magnetic field, the MDDI is predicted to induce the Einstein-de Haas effect\cite{PhysRevLett.96.080405,PhysRevLett.96.190404,PhysRevLett.99.130401,PhysRevLett.106.140403}.

In this paper, we theoretically show that the MDDI in a spinor BEC induces the spin Hall effect. In a spinor system, the geometric Hall effect has been observed \cite{PhysRevLett.111.245301}, where the spin-gauge symmetry of a ferromagnetic BEC creates a Lorentz-like force from a skyrmionic spin texture. Here, we consider a spin-1 BEC in the magnetic sublevel $m=0$ (the polar state), which preserves the time-reversal symmetry, and calculate the spin Hall conductivity due to the thermally and quantum-mechanically excited atoms in the magnetic sublevels $m=\pm 1$ using the Bogoliubov approximation. Though we calculate for a spin-1 BEC for simplicity, our calculation can be applicable for larger spins as far as the $m=0$ condensate is stable. Since the MDDI conserves the total, i.e., spin plus orbital, angular momentum of two colliding atoms, the orbital angular momenta of the atoms excited in the $m=1$ and $-1$ states differ by $2\hbar$. This is the origin of the spin Hall effect. According to the linear response theory, the spin Hall conductivity is calculated from the off-diagonal correlation of the mass- and spin-current operators, which are defined from the equations of continuity. We find that the corrections of the current operators due to the MDDI are essential for the emergence of the spin Hall effect.

The rest of the paper is organized as follows. In Sec. II, we introduce the Bogoliubov Hamiltonian for the polar state of a spin-1 dipolar BEC and discuss the properties of the excitations. In Sec. III, the mass- and spin-current operators are defined from the continuity equations. The time derivative of the mass-current operator shows that the MDDI causes Lorentz-like forces that work in the opposite directions for atoms excited in the $m=1$ and $-1$ states. In Sec. IV, the emergence of the spin Hall effect is demonstrated through evaluating the spin Hall conductivity using the linear response theory. We discuss the temperature and the quadratic Zeeman energy dependence of the spin Hall conductivity, and evaluate it for a realistic situation. Section V concludes the paper. The detailed calculation of the spin Hall conductivity is given in Appendix A.
\section{Bogoliubov Hamiltonian}
We consider a system of spin-1 Bose atoms. For the sake of simplicity, we neglect the confining potential and assume a spatially uniform system with volume $\Omega$.
The Hamiltonian of this system is given by
\begin{equation}
\hat H=\sum_{\mathbf k,m}(\epsilon_{\mathbf k}+q_{\rm Z} m^{2})\hat{a}^{\dagger}_{\mathbf k,m}\hat{a}_{\mathbf k,m}+\hat{H}_{\rm s}+\hat{H}_{\rm dd},\label{eq:Hamiltonian}
\end{equation}
where $\hat{a}_{\mathbf k,m}$ and $\hat{a}^{\dagger}_{\mathbf k,m}$ are the annihilation and creation operators for spin-1 atoms with the momentum $\bf k$ in the magnetic sublevel $m=1, 0,$ and $-1$, $\epsilon_{\mathbf k}=k^{2}/2M $ with $M$ being the atomic mass, and $q_{\rm Z}$ is the quadratic Zeeman energy per atom. Here, we assume that the external magnetic field is absent and $q_{\rm Z}>0$ is tuned by applying a microwave field\cite{PhysRevA.73.041602,PhysRevA.75.053606}. The short-range part of the interaction $\hat{H}_{\rm s}$ is divided into the density-density interaction and the spin-exchange interaction and given by
\begin{eqnarray}
&&\hat H_{\rm s}=\frac{1}{2\Omega}\sum_{\scriptscriptstyle{m_{1}m_{2}m_{3}m_{4}}}\sum_{\scriptscriptstyle{\mathbf k_{1}\mathbf k_{2}\mathbf k_{3}\mathbf k_{4}}}\{ c_{0}\delta_{m_{1}m_{4}}\delta_{m_{2}m_{3}}\nonumber \\
&+&\displaystyle{c_{1} (\mathbf F)_{\scriptscriptstyle{m_{1}m_{4}}} \cdot (\mathbf F)_{\scriptscriptstyle{m_{2}m_{3}}}\}
\delta_{\scriptscriptstyle{\mathbf k_{1}+\mathbf k_{2},\mathbf k_{3}+\mathbf k_{4}}}\hat{a}^{\dagger}_{\scriptscriptstyle{\mathbf k_{1}m_{1}}}\hat{a}^{\dagger}_{\scriptscriptstyle{\mathbf k_{2}m_{2}}}\hat{a}_{\scriptscriptstyle{\mathbf k_{3}m_{3}}}\hat{a}_{\scriptscriptstyle{\mathbf k_{4}m_{4}}}},\nonumber \\
\end{eqnarray}
where $\mathbf F$ is the vector of the spin-1 matrices, and the interaction coefficients are given by $c_{0}=4\pi (2a_{2}+a_{0})/3M$ and $c_{1}=4\pi (a_{2}-a_{0})/3M$ with $a_\mathcal{F}$ being the s-wave scattering length of two colliding atoms with total spin $\mathcal{F}=0 {\ \rm and\ } 2$. The sign of $c_1$ determines the magnetism of the ground state: The condensate is ferromagnetic for $c_1<0$ and polar (or anti-ferromagnetic) for $c_1>0$ \cite{doi:10.1143/JPSJ.67.1822,PhysRevLett.81.742}. The spin-1 $^{87}$Rb atoms and the spin-1 $^{23}$Na atoms are known to be ferromagnetic and polar, respectively.
The MDDI Hamiltonian is given by
\begin{eqnarray}
&&\hat H_{\rm dd}\nonumber \\
&=&\frac{c_{\rm dd}}{2\Omega}\sum_{\scriptscriptstyle{m_{1}m_{2}m_{3}m_{4}}}\sum_{\scriptscriptstyle{\mathbf k_{1}\mathbf k_{2}\mathbf k_{3}\mathbf k_{4}}}\sum_{\nu \nu^{'}}\tilde Q_{\nu \nu^{'}}(\mathbf k_{4}-\mathbf k_{1})\nonumber \\
&&(\mathbf F_{\nu})_{\scriptscriptstyle{m_{1}m_{4}}} (\mathbf F_{\nu^{'}})_{\scriptscriptstyle{m_{2}m_{3}}}\delta_{\scriptscriptstyle{\mathbf k_{1}+\mathbf k_{2},\mathbf k_{3}+\mathbf k_{4}}} \hat{a}^{\dagger}_{\scriptscriptstyle{\mathbf k_{1}m_{1}}}  \hat{a}^{\dagger}_{\scriptscriptstyle{\mathbf k_{2}m_{2}}} \hat{a}_{\scriptscriptstyle{\mathbf k_{3}m_{3}}}\hat{a}_{\scriptscriptstyle{\mathbf k_{4}m_{4}}},\nonumber \\ \label{eq:Hdd}
\end{eqnarray}
where $c_{\rm dd}=\mu_{0}g_{F}^{2}\mu_{\rm B}^{2}/(4\pi)$ with $\mu_{0}$ being the magnetic
permeability of the vacuum, $\mu_{\rm B}$ the Bohr magneton, and $g_{F}$ the Lande's hyperfine g-factor. For the case of spin-1 $^{87}$Rb atoms and spin-1 $^{23}$Na atoms, which have nuclear spin 3/2 and electron spin 1/2, we have $g_F=1/2$. The kernel of the MDDI $\tilde Q_{\nu \nu^{'}}(\mathbf k)$ is given by
\begin{equation}
\tilde Q_{\nu \nu^{'}}(\mathbf k)=-\frac{4\pi}{3}(\delta_{\nu \nu'}-3\hat k_{\nu}\hat k_{\nu^{'}}),\label{eq:kernel}
\end{equation}
where $\hat{{\bf k}}={\bf k}/|{\bf k}|$. $\tilde{Q}_{\nu\nu'}({\bf k})$ is the Fourier transform of the MDDI kernel in the real space
\begin{equation}
Q_{\nu\nu'}({\bf r})=\frac{\delta_{\nu\nu'}-3\hat{r}_\nu\hat{r}_{\nu'}}{r^3},
\end{equation}
where $\bf r$ is the relative position of two dipole moments, $r=|{\bf r}|$, and $\hat{\bf r}={\bf r}/r$.
From Eqs. (\ref{eq:Hdd}) and (\ref{eq:kernel}),  one can see that the spin and orbital degrees of freedom are coupled, and in this sense, the MDDI is regarded as an SOI. It is well known that in solid-state ferromagnets, the MDDI induces spatially varying magnetic structures, such as spin vortices and magnetic bubbles \cite{hubert2008magnetic}. Similar structures are predicted to appear in ferromagnetic BECs \cite{PhysRevLett.97.020401,PhysRevLett.97.130404}. On the other hand, for polar BECs, non-uniform magnetic structures appear only when condensates are highly oblate or the MDDI is stronger than the spin-exchange interaction \cite{PhysRevLett.97.020401}. This is because the MDDI is the interaction between local magnetizations and the weak MDDI does not contribute to the ground state in the mean-field level when the condensate has no spontaneous magnetization. Hence, the leading contribution of the MDDI to the polar BEC is in the excitation spectrum.

In the following calculation, we consider a polar BEC $(c_1>0)$ and assume that the atoms are condensed in the $({\bf k}, m)=({\bf 0},0)$ state. This is the ground state for weak MDDIs. The stability of the ground state shall be confirmed by the excitation spectrum obtained below. Following the Bogoliubov theory, which describes excitations at low temperature, we expand the Hamiltonian (1) up to the second order in fluctuations  [i.e., $\hat{a}_{{\bf k},m}$ and $\hat{a}_{{\bf k},m}^\dagger$ with $({\bf k}, m)\neq ({\bf 0},0)$], obtaining the Bogoliubov Hamiltonian
\begin{eqnarray}
\hat H_{\rm B}&=&\sum_{\mathbf k\ne0}(\epsilon_{\mathbf k} + c_{0}n)a^{\dagger}_{\mathbf k, 0}a_{\mathbf k, 0} \nonumber \\
&+&\frac{1}{2}\sum_{\mathbf k\ne0}c_{0}n(a_{\mathbf k, 0}a_{\mathbf k, 0}+a_{\mathbf k, 0}^{\dagger}a_{\mathbf k, 0}^{\dagger})\nonumber \\
&+&\frac{1}{2}\sum_{\mathbf k}\vec{\hat \psi^{\dagger}_{\mathbf{k}}}\mathbf{H}_{\mathbf{k}}\vec{\hat \psi_{\mathbf{k}}}, \label{eq:HB}
\end{eqnarray}
where $n$ is the number density of the condensate. Here, the $m=0$ component and the other two components are decoupled from each other, and their contributions to $\hat{H}_{\rm B}$ are divided into the first two terms and the last term in the right-hand side of Eq. (\ref{eq:HB}). The former is not affected by the MDDI and results in the conventional phonon spectrum, $E_{{\bf k},0} = \sqrt{\epsilon_k(\epsilon_k+2c_0n)}$. Since the $m=0$ component is not related to the spin current, whose definition will be given in the next section, our main concern is the last term of Eq. (\ref{eq:HB}). Here, $\vec{\hat{\psi}}_{\bf k}^\dagger$ and $\vec{\hat{\psi}}_{\bf k}$ are vectors of the creation and annihilation operators:
$\vec{\hat{\psi}^{\dagger}}_{\mathbf k}=(a^{\dagger}_{\mathbf k, \uparrow},a^{\dagger}_{\mathbf k, \downarrow},a_{-\mathbf k, \uparrow}, a_{-\mathbf k, \downarrow})$
 and 
$\vec{\hat{\psi}}_{\mathbf k}=(a_{\mathbf k, \uparrow}, a_{\mathbf k, \downarrow},a^{\dagger}_{-\mathbf k, \uparrow},a^{\dagger}_{-\mathbf k, \downarrow})^{\rm T}$, where T denotes the transpose, and here and hereafter we use $m=\uparrow$ and $\downarrow$ instead of $m=1$ and $-1$. The $4\times 4$ matrix ${\bf H}_{\bf k}$ is generally written using submatrices ${\bf f}_{\bf k}$ and ${\bf g}_{\bf k}$ as
\begin{equation}
\mathbf{H}_{\mathbf{k}}=
\left(
\begin{array}{cc}
\mathbf f_{\mathbf k}& \mathbf g_{\mathbf k} \\
\mathbf g_{-\mathbf k}^{\ast}& \mathbf f_{-\mathbf k}^{\ast}
\end{array}
\right).
\end{equation}
In the present case, $\mathbf f_{\mathbf{k}}$ and $\mathbf g_{\mathbf{k}}$ are respectively given by
\begin{equation}
\mathbf f_{\mathbf{k}}=
\left(
\begin{array}{ccc}
\epsilon_{\mathbf k} + c_{1}n + q_{\rm Z} +d_{5,\mathbf k}& d_{-,\mathbf k} \\
d_{+,\mathbf k}& \epsilon_{\mathbf k} + c_{1}n + q_{\rm Z}+d_{5,\mathbf k}
\end{array}
\right),
\end{equation}
\begin{equation}
\mathbf g_{\mathbf{k}}=
\left(
\begin{array}{ccc}
d_{-,\mathbf k}& c_{1}n + d_{5,\mathbf k} \\
c_{1}n + d_{5,\mathbf k}& d_{+,\mathbf k}
\end{array}
\right),
\end{equation}
where
\begin{eqnarray}
d_{\pm,\bf k} = \frac{2\pi c_{dd}n}{k^{2}} (k_x \pm i k_y)^2
,\nonumber
\end{eqnarray}
\begin{eqnarray}
d_{5,\mathbf k}=\frac{2\pi c_{dd}n}{3k^{2}}(k^{2}-3k_{z}^{2})
\end{eqnarray}
describe the contribution of the MDDI.

We diagonalize $\mathbf H_{\mathbf k}$ by introducing the Bogoliubov transformation $\vec{\hat{\psi}}_{\mathbf k}=\mathbf{T}_{\mathbf k} \vec{\hat{\phi}}_{\mathbf k}$, where $\vec{\hat{\phi}}_{\mathbf k}=(b^{\dagger}_{\mathbf k, \uparrow},b^{\dagger}_{\mathbf k, \downarrow},b_{-\mathbf k, \uparrow}, b_{-\mathbf k, \downarrow})^{\rm T}$ with $b^{\dagger}_{\mathbf k, \downarrow}$ $(b_{\mathbf k, \downarrow})$ being the creation (annihilation) operator of quasi particles. To ensure the bosonic commutation relations between quasiparticles, $\mathbf{T}_{\mathbf k}$ is a $4\times 4$ pseudo-unitary matrix satisfying $\mathbf{T}_{\mathbf k}^\dagger \sigma_{3} \mathbf{T}_{\mathbf k} = \sigma_{3}$, where $\sigma_3$ is given by
\begin{equation}
\sigma_{3}=
\left(
\begin{array}{cc}
\mathbf{1}_{2\times2}& 0 \\
0& -{\mathbf 1}_{2\times2}
\end{array}
\right),
\end{equation}
with ${\bf 1}_{2\times 2}$ being a $2\times 2$ identity matrix.
Rewriting $\mathbf{T}_{\bf k}$ as
\begin{equation}
\mathbf T_{\mathbf k}=
\pmatrix{
\mathbf U_{\mathbf k} & \mathbf V^{\ast}_{-\mathbf k} \cr
\mathbf V_{\mathbf k} & \mathbf U^{\ast}_{-\mathbf k} \cr
}
\end{equation}
and solving the Bogoliubov equation,
\begin{eqnarray}
\mathbf{T}_{\bf k}^\dagger H_{\bf k}\mathbf{T}_{\bf k}={\rm Diag}[E_{{\bf k},\uparrow},E_{{\bf k},\downarrow},E_
{-{\bf k},\uparrow},E_{-{\bf k},\downarrow}],
\end{eqnarray}
we obtain the energy spectra
\begin{eqnarray}
E_{{\bf k},m} = \sqrt{(\epsilon_k+q_{\rm Z})(\epsilon_k+q_{\rm Z}+2 c_m n)},\label{eq:energy}
\end{eqnarray}
where $m=\uparrow$ and $\downarrow$, and we define
\begin{eqnarray}
c_\uparrow &=& c_1 + \frac{4\pi c_{\rm dd}}{3}(2-3\cos^2\theta),\\
c_\downarrow &=& c_1 - \frac{4\pi c_{\rm dd}}{3},
\end{eqnarray}
and $\theta = \tan^{-1}(\sqrt{k_x^2+k_y^2}/k_z)$.
In the absence of the MDDI, these two modes are degenerate ($c_\uparrow=c_\downarrow$). In particular when $q_{\rm Z}=0$, these modes become gapless magnon modes, which are the Nambu-Goldstone modes associated with the spontaneous breaking of the spin-rotation symmetry in the ground state.  Since the excitations induce local magnetizations, which then interact with each other via the MDDI, the MDDI lifts the degeneracy of two magnons. For $c_{\rm dd}>3[c_1+q_{\rm Z}/(2n)]/(4\pi)$,  the energy spectra become pure imaginary at long wavelengths and the system becomes dynamically unstable. In other words, the condensate in the $m=0$ state is stable for  $c_{\rm dd}<3[c_1+q_{\rm Z}/(2n)]/(4\pi)$.

The role of the MDDI becomes clearer when we see the corresponding eigenmodes, which are given by
\begin{eqnarray}
\mathbf U_{\mathbf k}&=&
\pmatrix{
\mathbf u_{\mathbf k,\uparrow} & \mathbf u_{\mathbf k,\downarrow}
}=
\pmatrix{
-e^{-i \phi}u_{\mathbf k,\uparrow} & -e^{-i \phi}u_{\mathbf k,\downarrow} \cr
-e^{i \phi}u_{\mathbf k,\uparrow} & e^{i \phi}u_{\mathbf k,\downarrow} \cr
}, \nonumber \\
\mathbf V_{\mathbf k}&=&
\pmatrix{
\mathbf v_{\mathbf k,\uparrow} & \mathbf v_{\mathbf k,\downarrow}
}=
\pmatrix{
e^{i \phi}v_{\mathbf k,\uparrow} & -e^{i \phi}v_{\mathbf k,\downarrow} \cr
e^{-i \phi}v_{\mathbf k,\uparrow} & e^{-i \phi}v_{\mathbf k,\downarrow} \cr
},\nonumber \\ \label{eq:Bog_T}
\end{eqnarray}
with $\phi=\rm tan^{-1}(\it{k_{y}}/\it{k_{x}})$
and
\begin{equation}
u_{\mathbf k,m}=\sqrt{\frac{\epsilon_{k}+q_{\rm z}+c_{m}n+E_{\mathbf k,m}}{2E_{\mathbf k,m}}},
\end{equation}
\begin{equation}
v_{\mathbf k,m}=\sqrt{\frac{\epsilon_{k}+q_{\rm z}+c_{m}n-E_{\mathbf k,m}}{2E_{\mathbf k,m}}}.
\end{equation}

Note here that the eigenmodes have the spin-dependent phase factor $e^{\pm i \phi}$. This is the consequence of the spin-momentum locking due to the MDDI. To see this, we move to the mean-field description. In the presence of the quasiparticle $\hat{b}_{{\bf k}, \uparrow}$, for example, the mean-field order parameter is given by
\begin{eqnarray}
\bf \Psi_{{\bf k},\uparrow} \equiv
\left(
\begin{array}{c}
\Psi_1 \\ 
\Psi_0 \\ 
\Psi_{-1} 
\end{array}\right)=
\left(
\begin{array}{c}
0 \\
\sqrt{n} \\ 
0 
\end{array}\right)\nonumber &-&
\left(
    \begin{array}{c}
      u_{\bf k,\uparrow}e^{-i\phi} \\
      0 \\
      u_{\bf k,\uparrow}e^{i\phi} \\
    \end{array}
  \right)e^{i(\bf{k}\cdot \bf{r}-E_{\bf{k},\uparrow}t)}\nonumber \\
&+&\left(
    \begin{array}{c}
      v_{\bf{k},\uparrow}e^{-i\phi} \\
      0 \\
      v_{\bf{k},\uparrow}e^{i\phi} \\
    \end{array}
  \right)e^{-i(\bf{k}\cdot \bf{r}-E_{\bf{k},\uparrow}t)},\nonumber \\
\end{eqnarray}
from which the spin expectation value is calculated to be
\begin{eqnarray}
\langle {\bf F}\rangle_{{\bf k},\uparrow} &\equiv& {\bf \Psi}_{{\bf k},\uparrow}^\dagger {\bf F} {\bf \Psi}_{{\bf k},\uparrow}\nonumber \\
&=&-2\sqrt{2}(u_{\mathbf{k},\uparrow}-v_{\mathbf{k},\uparrow})\cos(\mathbf{k}\cdot \mathbf{r} - E_{\mathbf{k},\uparrow}t) \left(
    \begin{array}{c}
      \cos{\phi} \\
      \sin{\phi} \\
      0 \\
    \end{array}
  \right).\nonumber \\
\end{eqnarray}
Namely, the $\hat{b}_{{\bf k},\uparrow}$ mode describes the spin density wave in which the spin orientation is parallel to the projected momentum vector onto the $xy$ plane ${\bf k}_{\perp}\equiv{\bf k}-({\bf k}\cdot \hat{z})\hat{z}$. In a similar manner, the spin expectation value for the $\hat{b}_{{\bf k},\downarrow}$ mode is given by
\begin{eqnarray}
\langle {\bf F}\rangle_{{\bf k},\downarrow} =2\sqrt{2} (u_{k,\downarrow}-v_{k,\downarrow})\sin(\mathbf{k}\cdot \mathbf{r} - E_{\mathbf{k},\uparrow}t) \left(
    \begin{array}{c}
      - \sin{\phi} \\
      \cos{\phi} \\
      0 \\
    \end{array}
  \right).\nonumber \\
\end{eqnarray}
In this case, the spin orientation is perpendicular to ${\bf k}_\perp$. Figure 1 shows the schematic picture of the spin density waves. The spin configuration for the $\downarrow$ mode always has negative MDDI energy, whereas the MDDI energy for the $\uparrow$ mode depends on the angle of ${\bf k}$ from the $z$ axis. This is the underlying physics of the energy spectra in Eq. (\ref{eq:energy}). In both cases of the $\uparrow$ and $\downarrow$ modes, the spin orientation relative to the momentum vector is locked, and this spin-momentum locking results in the nonzero spin Hall conductivity, which we will show in Sec. IV.
\section{Mass-current and spin-current operators}
Before calculating the spin Hall conductivity, we need to define the mass-current and spin-current operators. Since the atoms in the $({\bf k}, m)$ state conveys momentum $\hbar{\bf k}$ and spin $m\hbar$, the naive definitions of the mass-current and spin-current operators are given by $\sum_{\bf k}\sum_{m=0,\pm1}\hbar\mathbf{k}/M \hat{a}_{{\mathbf{k}},m}^\dagger\hat{a}_{\mathbf{k},m}$ and $\sum_{\bf k}\sum_{m=0,\pm1}m\hbar\mathbf{k}/M \hat{a}_{{\mathbf{k}},m}^\dagger\hat{a}_{\mathbf{k},m}$, respectively. This is correct in the absence of the MDDI. However, in the presence of the MDDI, because of the momentum dependence of the MDDI Hamiltonian the current operators should be modified so as to satisfy the equation of continuity.

Here, we start from the Fourier component of the number-density operator and that of the $z$ component of the spin-density operator
\begin{equation}
\hat{\rho}^{\rm M,S}(\mathbf q)=\sum_{\bf k}\sum_{m=\uparrow,\downarrow}X^{\rm M,S}_{m}\hat{a}^{\dagger}_{\mathbf k + \frac{\mathbf q}{2},m}\hat{a}_{\mathbf k - \frac{\mathbf q}{2},m},
\end{equation}
where $X^{\rm M}_{m}=1$ and $X^{\rm S}_{m}=m$, and 
the superscripts M and S denote mass and spin, respectively. Since the phonon and magnon excitations are decoupled, we omit the contribution from the $m=0$ component to the number-density operator.
The corresponding current operators $\hat{ \mathbf J}_{\mathbf q}^{\rm M,S}$ are defined from the Fourier transform of the equation of continuity:
\begin{equation}
\frac{\partial \hat{\rho}^{\rm M}(\mathbf q)}{\partial t}
= \frac{1}{i \hbar}[\hat{\rho}^{\rm M} (\mathbf q),\hat{H}]=-i\mathbf q \cdot \hat{ \mathbf J}_{\mathbf q}^{\rm M},\label{eq:mass-cnsrv}
\end{equation}
\begin{equation}
\frac{\partial\hat{\rho}^{\rm S}
(\mathbf q)}{\partial t} = \frac{1}{i \hbar}[\hat{\rho}^{\rm S} (\mathbf q),\hat{H}]=-i\mathbf q \cdot \hat{ \mathbf J}_{\mathbf q}^{\rm S}+T.\label{eq:spin-cnsrv}
\end{equation}
Note that whereas the number of atoms is always a good quantum number of the Hamiltonian (\ref{eq:Hamiltonian}), the $z$ component of the spin angular momentum is not conserved in the presence of the MDDI. This is because the MDDI couples the spin and orbital angular momenta of atoms, and the angular-momentum-transfer process works as source and drain of spin in the equation of continuity. $T$ in Eq. (\ref{eq:spin-cnsrv}) represents such terms. We calculate the current operators in the long wavelength limit (${\bf q}\to{\bf 0}$) by approximating $\hat{H}$ with $\hat{H}_{\rm B}$ and expanding the commutators in Eqs. (\ref{eq:mass-cnsrv}) and (\ref{eq:spin-cnsrv}) with respect to ${\bf q}$. The obtained results are given by

\begin{equation}
\hat{\bf{J}}^{\rm M,S}_{\mathbf q \to 0}=\sum_{\mathbf k} \vec{\hat{\psi}}^{\dagger}_{\mathbf k}\bf{J}^{\rm M,S}_{\mathbf k} \vec{\hat{\psi}}_{\mathbf k},\label{eq:JMS}
\end{equation}
\begin{equation}
\bf{J}^{\rm M}_{\mathbf k}=
\left(
\begin{array}{cccc}
\frac{1}{\hbar}\frac{\partial \mathbf f_{\mathbf k}}{\partial \mathbf k}& 0 \\
0& \frac{1}{\hbar}\frac{\partial \mathbf f^{*}_{\mathbf -k}}{\partial \mathbf k}
\end{array}
\right),\label{eq:JM}
\end{equation}
\begin{equation}
\mathbf{J}^{\rm S}_{\mathbf k}=
\left(
\begin{array}{cccc}
\sigma_{z}& 0 \\
0& \sigma_{z}
\end{array}
\right)\mathbf{J}^{\rm M}_{\mathbf k}+\frac{i}{\hbar} \frac{\partial d_{5,\mathbf k}}{\partial \mathbf k}
\left(
\begin{array}{cccc}
0& \sigma_{y} \\
-\sigma_{y}& 0
\end{array}
\right),\label{eq:JS}
\end{equation}
where $\sigma_i (i=x,y,z)$ are the Pauli matrices given by
\begin{eqnarray}
\sigma_{x}=
\pmatrix{
0 & 1 \cr
1 & 0 \cr
},
\sigma_{y}=
\pmatrix{
0 & -i \cr
i & 0 \cr
},
\sigma_{z}=
\pmatrix{
1 & 0 \cr
0 & -1 \cr
}.\nonumber \\
\end{eqnarray}

The correction to the mass-current operator due to the MDDI includes not only the diagonal terms ($\frac{\partial d_{5,{\bf k}}}{\partial {\bf k}}\hat{a}_{{\bf k},m}^\dagger \hat{a}_{{\bf k},m}$) but also the off-diagonal terms ($\frac{\partial d_{+,{\bf k}}}{\partial {\bf k}} \hat{a}_{{\bf k},\uparrow}^\dagger\hat{a}_{{\bf k},\downarrow}$ and its Hermite conjugate). The former is the correction of the dispersion due to the MDDI, whereas the latter describe the process that atoms convey a fixed momentum with changing their spin state. The same terms also appear in the spin-current operator, but the off-diagonal ones are cancel out by taking the summation with respect to ${\bf k}$ in Eq. (\ref{eq:JMS}), which is consistent with our intuition. The spin-current operator further includes the paring terms ($\frac{\partial d_{5,{\bf k}}}{\partial {\bf k}} \hat{a}_{{\bf k},\uparrow}^\dagger\hat{a}_{-{\bf k},\downarrow}^\dagger$ and its Hermite conjugate), which come from the fact that
the atoms in the $({\bf k},\uparrow)$ state and those in the $(-{\bf k},\downarrow)$ state convey the spin current in the same direction. 

Here we comment that the torque term $T$ in Eq. (\ref{eq:spin-cnsrv}) does not contribute to the spin Hall effect. In the calculation of the spin Hall conductivity, we evaluate the correlation function between $\hat{\bf J}_{{\bf q}\to \bf 0}^{\rm M}$ and $\hat{\bf J}_{{\bf q}\to\bf 0}^{\rm S}$ (see next section). When we consider the correlation between $\hat{\bf J}_{{\bf q}\to\bf 0}^{\rm M}$ and $T$, it always vanishes because the matrix elements of $T$ in the Nambu spinor basis consists of even functions of ${\bf k}$, $d_{\pm}({\bf k})$, whereas those of $\hat{\bf J}_{{\bf q}\to\bf 0}^{\rm M}$ consists of odd functions of ${\bf k}$. This result means that the DC mass current cannot directly cause time derivative of the spin density, and vice versa. 
\begin{figure}[ht]
\includegraphics[width=65mm]{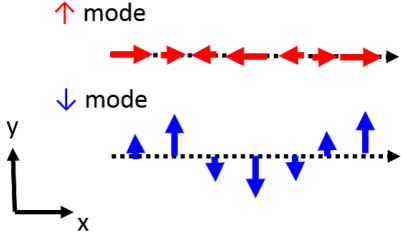}
\caption{
(color online). Schematic picture of spin density wave in the $\uparrow/\downarrow$ modes propagating along the $x$ direction. In the $\uparrow$ mode ($\downarrow$ mode), the spin density arises parallel (perpendicular) to the momentum ${\bf k} \parallel \hat{x}$.}
\end{figure}

An intuitive interpretation for the appearance of the spin Hall effect is obtained by taking the time derivative of the mass-current operator [Eqs. (\ref{eq:JMS}) and (\ref{eq:JM})]:
\begin{eqnarray}
\langle \dot{\hat{\mathbf J}}^{\rm M}_{\mathbf q \to 0} \rangle &=& \langle \frac{1}{i\hbar}[\hat{\mathbf J}^{\rm M}_{\mathbf q \to 0},\hat{H}_{B}] \rangle \nonumber \\
&=& \frac{1}{i\hbar}\sum_{\mathbf k} \left( \frac{\partial \mathbf f_{\mathbf k}}{\partial \mathbf k}\mathbf f_{\mathbf k}-\mathbf f_{\mathbf k}\frac{\partial \mathbf f_{\mathbf k}}{\partial \mathbf k}\right)_{mm'}  \langle \hat{a}^{\dagger}_{\mathbf k, m}\hat{a}_{\mathbf k, m'} \rangle \nonumber \\
&=& \sum_{\mathbf k}\left[
F_{+}({\bf k}) \langle \hat{a}^{\dagger}_{\mathbf k, \uparrow}\hat{a}_{\mathbf k, \uparrow} \rangle + F_{-}({\bf k}) \langle \hat{a}^{\dagger}_{\mathbf k, \downarrow}\hat{a}_{\mathbf k, \downarrow} \rangle \right], \nonumber \\ \label{eq:eom}
\end{eqnarray}
where $\langle...\rangle$ means the expectation value under an eigenstate of the Bogoliubov Hamiltonian $\hat{H}_{\rm B}$, and
\begin{eqnarray}
F_{+}({\bf k}) &\equiv& \frac{1}{i\hbar} \left(
\frac{\partial \mathbf f_{\mathbf k}}{\partial \mathbf k}\mathbf f_{\mathbf k}-\mathbf f_{\mathbf k}\frac{\partial \mathbf f_{\mathbf k}}{\partial \mathbf k}
\right)_{\uparrow\uparrow}\nonumber \\
&=&
-\frac{(4\pi c_{{\rm dd}}n)^{2}}{\hbar}\frac{\partial \phi}{\partial {\bf k}},
\end{eqnarray}
\begin{eqnarray}
F_{-}({\bf k}) &\equiv& \frac{1}{i\hbar} \left(
\frac{\partial \mathbf f_{\mathbf k}}{\partial \mathbf k}\mathbf f_{\mathbf k}-\mathbf f_{\mathbf k}\frac{\partial \mathbf f_{\mathbf k}}{\partial \mathbf k}
\right)_{\downarrow\downarrow}\nonumber \\
&=&
+\frac{(4\pi c_{{\rm dd}}n)^{2}}{\hbar}\frac{\partial \phi}{\partial {\bf k}} = -F_{+}({\bf k}).
\end{eqnarray}
In Eq. (\ref{eq:eom}), the contributions from the terms including $\langle a a \rangle$ and $\langle a^{\dagger} a^{\dagger} \rangle$ cancel with each other. Although the remaining terms in the right-hand side of Eq. (\ref{eq:eom}) also cancel with each other, Eq. (\ref{eq:eom}) indicates that the excited atoms from the condensate are subjected to the spin-dependent Lorentz-like force ${\mathbf F}_{\pm}(\mathbf k)$ whose direction is perpendicular to ${\bf k}_{\perp}$ as shown in Fig. 2. This is the origin for the spin Hall effect.
\begin{figure}[h]
\begin{center}
\includegraphics[width=65mm]{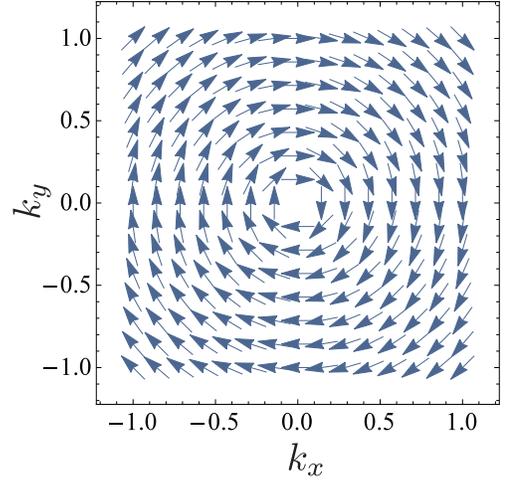}
\caption{(color online).The directions of spin-dependent Lorenz-like force $F_{+}({\bf k})$ acting on spin-up atoms in the momentum space.}\label{fig:force}
\end{center}
\end{figure}
\section{Spin Hall conductivity}
In this section, the spin Hall conductivity is calculated by using the linear response theory.
We first calculate the off-diagonal current-current response function $Q(i\omega)$ defined by
\begin{eqnarray}
&Q&(i\omega_{n})\nonumber \\
&=&\sum_{\bf k,k'}\int_{0}^{\beta}\langle {\rm T}_{\tau}\vec{\hat \psi^{\dagger}_{\mathbf{k}}}(\tau)\mathbf J_{\mathbf k,x}^{M}\vec{\hat \psi_{\mathbf{k}}}(\tau)\ \vec{\hat \psi^{\dagger}_{\mathbf{k'}}}\mathbf J_{\mathbf k',y}^{S}\vec{\hat \psi_{\mathbf{k'}}}\rangle e^{i\omega_{n}\tau}d\tau ,\nonumber \\
\label{eq:response}
\end{eqnarray}
where $\omega_{n}=2 \pi n / \beta$, $\beta=1/(k_{\rm B}T)$, $\rm{T}_{\tau}$ represents the ordering with respect to the imaginary time $\tau$, and the average $\langle ... \rangle$ is taken as $\langle ... \rangle \equiv \sum_{n}\langle n|e^{-\beta \hat{H}_{\rm{B}}}...|n\rangle /\sum_{n}\langle n|e^{-\beta \hat{H}_{\rm{B}}}|n\rangle$ with $|n\rangle$ being the eigenvector of the Bogoliubov Hamiltonian $\hat{H}_{\rm{B}}$ ($\hat{H}_{\rm{B}} |n\rangle=E_{n}|n\rangle$). Moving to the quasi-particle basis $\vec{\hat{\phi}}_{{\bf k}}$ and using $\langle b_{{\bf k},m}^{\dagger}b_{{\bf k},m} \rangle = n_{{\bf k},m} = 1/(e^{\beta E_{{\bf k},m}}-1)$, the response function is obtained as
\begin{eqnarray}
Q(i\omega_{n})&=&\sum_{\mathbf k}(n_{\mathbf k,\uparrow}-n_{\mathbf k,\downarrow}) \frac{4i \hbar \omega_{n}S(\mathbf k)}{-(i \hbar \omega_{n})^{2}+(E_{\mathbf k,\uparrow}-E_{\mathbf k,\downarrow})^{2}}  \nonumber \\
&+&\sum_{\mathbf k}(1+n_{\mathbf k,\uparrow}+n_{\mathbf k,\downarrow}) \frac{8i \hbar \omega_{n}P(\mathbf k)}{-(i \hbar \omega_{n})^{2}+(E_{\mathbf k,\uparrow}+E_{\mathbf k,\downarrow})^{2}} ,\label{eq:30} \nonumber \\
\end{eqnarray}
where
\begin{eqnarray}
S(&\mathbf k&)\equiv -i \frac{8 c_{\rm dd}n}{\hbar}(u_{\mathbf k,\uparrow}u_{\mathbf k,\downarrow}-v_{\mathbf k,\uparrow}v_{\mathbf k,\downarrow})\sin^{2}\theta \sin^{2}\phi \nonumber \\ 
&\biggl\{& \frac{8\pi c_{\rm dd}n}{3}\frac{\cos^{2}\theta}{k^{2}}(u_{\mathbf k,\uparrow}-v_{\mathbf k,\uparrow})(u_{\mathbf k,\downarrow}-v_{\mathbf k,\downarrow})\nonumber \\
&\quad&\quad\quad+2\frac{\hbar^{2}}{M}(u_{\mathbf k,\uparrow}u_{\mathbf k,\downarrow}+v_{\mathbf k,\uparrow}v_{\mathbf k,\downarrow})\biggr\},\label{eq:S}
\\
P(&\mathbf k&)\equiv -i \frac{8 c_{\rm dd}n}{\hbar}(u_{\mathbf k,\uparrow}v_{\mathbf k,\downarrow}-v_{\mathbf k,\uparrow}u_{\mathbf k,\downarrow})\sin^{2}\theta \sin^{2}\phi \nonumber \\ 
&\biggl\{& -\frac{8\pi c_{\rm dd}n}{3}\frac{\cos^{2}\theta}{k^{2}}(u_{\mathbf k,\uparrow}-v_{\mathbf k,\uparrow})(u_{\mathbf k,\downarrow}-v_{\mathbf k,\downarrow})\nonumber \\
&\quad&\quad\quad+2\frac{\hbar^{2}}{M}(u_{\mathbf k,\uparrow}v_{\mathbf k,\downarrow}+v_{\mathbf k,\uparrow}u_{\mathbf k,\downarrow})\biggr\}.\label{eq:P}
\end{eqnarray}
The detailed derivation of Eqs. (\ref{eq:30})-(\ref{eq:P}) is given in Appendix A.
The first term in Eq. (\ref{eq:30}) describes the contribution from the scattering between thermally excited Bogoliubov particles, while the second term comes from the virtual process of creation and annihilation of pairs of Bogoliubov particles.
The fact that $Q(i\omega_{n})$ becomes zero when $c_{\rm dd}=0$ indicates that the spin Hall effect in our system is purely induced by  the MDDI. 
 
To see the spin Hall effect, we apply a magnetic field gradient in the $y$ direction (${\bf B}=B' y \hat{z}$), which induces a spin current along the $y$ direction, and calculate the responding mass current in the $x$ direction. Namely, this is the inverse spin Hall effect. (The usual spin Hall effect also occurs when we apply a potential gradient; the spin Hall and inverse spin Hall effects are essentially the same.) The Hamiltonian of this perturbation is given by
\begin{eqnarray}
H^{'}&=&-g_{F}\mu_{{\rm B}} \int d\mathbf r B^{'}y \hat \rho^{S}(\mathbf r)\nonumber \\
&=& -g_{F}\mu_{{\rm B}} B^{'} \lim_{q_{y},\omega \to 0} \frac{\hat \rho^{S}(-q_{y})e^{-i \omega t}-\hat \rho^{S}(q_{y})e^{i \omega t}}{2iq_{y}}.\nonumber \\
\label{eq:mag}
\end{eqnarray}
According to the linear response theory, the spin Hall conductivity is derived as
\begin{eqnarray}
&\sigma^{\rm SH}_{xy}&\nonumber \\
&=&\frac{\langle J^{M}_{x} \rangle}{g_{F}\mu_{\rm B} B^{'}}=\lim_{\omega \to 0}\frac{Q^{\rm R}(\omega)-Q^{\rm R}(0)}{-i\omega} \\
&=& 4i
\sum_{\mathbf k} \left[ \frac{n_{\mathbf k,+}-n_{\mathbf k,-}}{(E_{\mathbf k,+}-E_{\mathbf k,-})^{2}} S(\mathbf k)+\frac{1+n_{\mathbf k,+}+n_{\mathbf k,-}}{(E_{\mathbf k,+}+E_{\mathbf k,-})^{2}} 2P(\mathbf k)
\right],\nonumber \\
\label{eq:conductivity}
\end{eqnarray}
where $Q^{\rm R}(\omega)$ is the retarded two-body Green's function given by analytical connection from Eq. (\ref{eq:response}).
The same conductivity also describes the response of the spin current to an applied potential gradient, i.e.,  $\langle J_{y}^{\rm S}\rangle=\sigma^{\rm SH}_{xy}V'$, for an applied potential $V({\bf r}) = V' x$.

We numerically evaluate the integral with respect to $\bf k$ in Eq. (\ref{eq:conductivity}), and calculate the dependence of $\sigma_{xy}^{\rm SH}$ on $c_{\rm dd}/c_{1}$ and $c_{1}n/(k_{\rm B}T)$. The result  for $q_{\rm Z}/(c_{1} n)=0.01$ is shown in Fig. 3, where the $\sigma_{xy}^{\rm SH}$ is scaled in units of $(\hbar\xi)^{-1}$ with $\xi \equiv \hbar/\sqrt{2M c_{1}n}$ being the spin healing length. In Fig. 3, the range of the vertical axis is $0\le c_{\rm dd}/c_{1}\le 3\left[2+q_Z/(c_{1}n)\right]/(8\pi)$ beyond which the polar BEC is no longer stable. As expected, the stronger MDDI leads to the larger $\sigma_{xy}^{\rm SH}$. The conductivity also becomes larger for higher temperature. This is because the current-current correlation in the present system comes from the correlation between excited atoms. We find that even at absolute zero, the spin Hall conductivity remains nonzero due to quantum fluctuations. For the case of spin-1 $^{23}$Na atoms, we have $c_{\rm dd}/c_{1}=1.3\times 10^{-2}$, where we have used the scattering lengths measured in Ref. \cite{PhysRevLett.99.070403}. The temperature dependence of the spin Hall conductivity of a $^{23}$Na BEC is depicted in Fig. 3(b).

Figure 4 shows the $q_{\rm Z}$ dependence of the conductivity, from which we see that the $\sigma_{xy}^{\rm SH}$ diverges as $q_{\rm Z}$ goes to zero. This is because the energy gaps of the magnon modes $\Delta_m=\sqrt{q_{\rm Z}(q_{\rm Z}+2c_{m} n)}\ (m=\uparrow, \downarrow)$ become zero at $q_Z=0$. Since there is no impurities in the present system, an infinitesimal perturbation can infinitely excite gapless magnon modes and the response to the perturbation diverges.  

\begin{figure}[h]
\subfigure{\includegraphics[width=80mm]{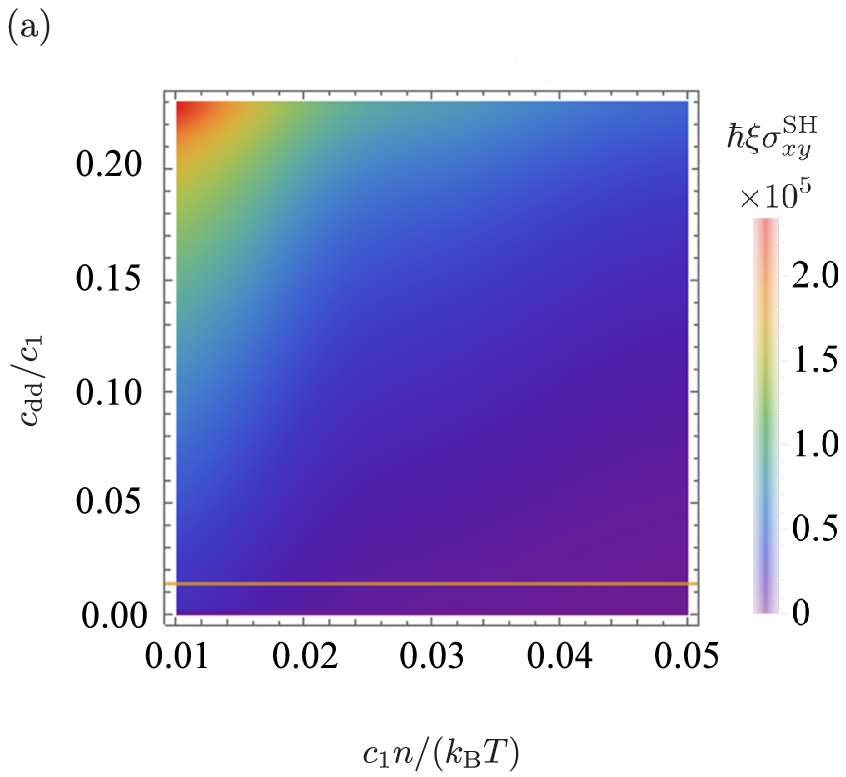}}
\subfigure{\includegraphics[width=75mm]{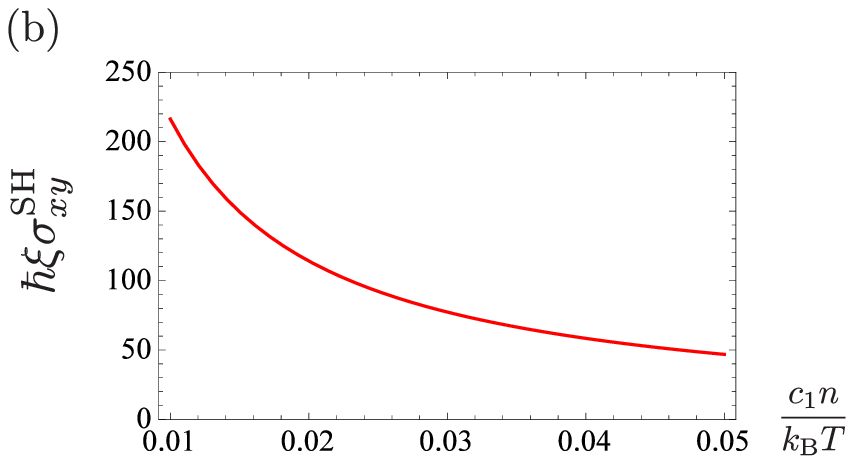}}
\caption{(color online). (a) Spin Hall conductivity $\sigma_{xy}^{\rm SH}$ as a function of temperature $c_{1}n/(k_{\rm B}T)$ and the MDDI strength $c_{\rm dd}/c_{1}$ at the quadratic Zeeman energy of $q_{\rm Z}/(c_{1} n)=0.01$, where $c_{1}$ is the strength of the spin-exchange interaction and $n$ is the number density of the system. The conductivity is scaled in units of $(\hbar\xi)^{-1}$ where $\xi\equiv \hbar/\sqrt{2M c_{1}1n}$ is the spin healing length of the condensate. The system of spin-1 $^{23}$Na atoms traces the yellow solid line in (a) at $c_{\rm dd}/c_{1}=1.3\times 10^{-2}$. (b) The same as (a) but for fixed $c_{\rm dd}/c_{1}=1.3\times 10^{-2}$.
}
\end{figure}
\begin{figure}[h]
\includegraphics[width=80mm]{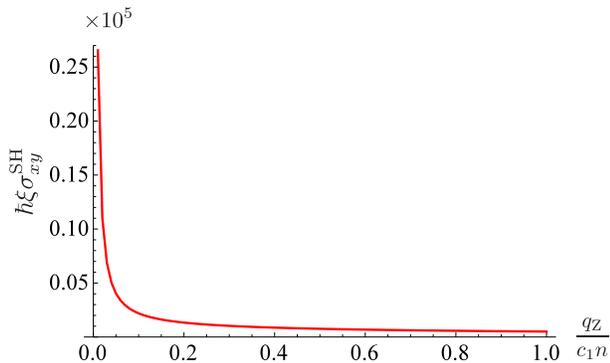}
\caption{(color online). The spin Hall conductivity $\sigma_{xy}^{\rm SH}$ as a function of the quadratic Zeeman energy $q_{\rm Z}/(c_{1}n)$ at $c_1n/(k_{\rm B}T)=0.01$ and $c_{\rm dd}/c_1=1.3\times 10^{-2}$. The conductivity diverges at $q_{\rm Z}=0$ because the magnon modes become gapless (see text). }
\end{figure}

For the realization of the inverse spin Hall effect in experiment, the strength of the external magnetic field should be controlled carefully, because the atomic spins rotate due to the linear Zeeman effect at the Lamor frequency $\hbar\omega_{\rm L}=g_{F}\mu_{\rm B}B$. When $\hbar\omega_{\rm L} \gg c_{\rm dd}n$, the effect of the MDDI is time-averaged over the Lamor precession period and the spin-orbit coupling effect is smeared out. So the external magnetic field gradient $B'$ must be prepared so as to satisfy
\begin{eqnarray}
g_{F} \mu_{\rm B}B' L < c_{\rm dd}n ,
\end{eqnarray}
where $L$ is the system size in the $y$ direction.
For the case of spin-1 $^{23}$Na atoms, which have the small MDDI of $c_{\rm dd}n\sim \hbar \times 4.7$~Hz for a characteristic number density $n\sim 2.3\times 10^{20}~{\rm m}^{-3}$, we obtain $B'< 1.0\times10^{-1}$~mG/mm for $L=10~\mu{\rm m}$.
Recent experimental technique makes it possible to control such a low magnetic field \cite{PhysRevLett.106.255303, PhysRevLett.108.045307}.

Figure 5 shows the transverse velocity of the excited atoms $\langle J^{\rm M}_{x}\rangle/N^{\rm ex}$ at various temperatures in response to the magnetic field gradient of $B'=1.0\times 10^{-2}$~mG/mm, where $N^{\rm ex}$ denotes the number of atoms excited in the $m=\pm 1$ states. Here, we have used the mass of the $^{23}$Na atom and the number density $n=2.3\times 10^{20}~{\rm m}^{-3}$. Because $N^{\rm ex}$ becomes larger for smaller $c_{1}n/(k_{\rm B}T)$, the mean velocity $\langle J^{\rm M}_{x} \rangle/N^{\rm ex}$ becomes smaller as $c_{1}n/(k_{\rm B}T)$ becomes smaller. As we can see from Fig. 5, the transverse velocity becomes up to a few $\mu$m at small $q_{\rm Z}$. For example, at $q_{\rm Z}=\hbar \times 1$~Hz and $c_1n/(k_{\rm B}T)=3.5\times10^{-3}$, which correspond to $T/T_{{\rm BEC}}\sim0.3$ and $N^{\rm ex}/N\sim 0.3$, the transverse velocity becomes $6.1~\mu$m/s. Suppose that the size of the BEC in the $x$ direction is $10~\mu$m. Then, most of the excited atoms are accumulated in the positive $x$ region within a second, which may be observed in experiments, although the effect of the finite size as well as the trapping potential should be included for a quantitative discussion.

So far, we have discussed the spin Hall effect in a spin-1 BEC. The same argument is applicable for larger spin systems as far as the condensate in the $m=0$ state is stable. In the case of a BEC of $^{52}$Cr atoms in the $m=0$ state, for example, the excitations in the $m=\pm1$ states, $m=\pm 2$ states, and $m=\pm 3$ states are decoupled from each other. Though the last one is dynamically unstable\cite{PhysRevLett.96.190404}, the instability can be removed by applying a Laser induced quadratic Zeeman effect\cite{PhysRevA.75.053606}. In this case, the effective interaction between the condensate and the $m=\pm1$ components is given by $\tilde{c}_1=4\pi \hbar^{2} (25a_{6}-3a_{4}-11a_{2}-11a_{0})/(77M)$ where the scattering lengths are given in Refs. \cite{PhysRevLett.94.183201,private}. On the other hand, since the g-factor for the $^{52}$Cr atom is 2 and the ${\bf F}$ in Eq. (3) becomes a vector of the spin-3 matrices, $d_{\pm,{\bf k}}$ and $d_{5,{\bf k}}$ defined in Eq. (9) are multiplied by 96, i.e., the effective MDDI for $^{52}$Cr atoms is given by $\tilde{c}_{\rm dd}=96c_{\rm dd}$. As a result, we have $\tilde{c}_{\rm dd}/\tilde{c}_{1}=0.08$ for $^{52}$Cr atoms, and hence the higher spin Hall conductivity is expected.

We also comment that although we have calculated within the Bogoliubov approximation, the motion of the normal component will expected to excite the motion of the condensate. For example,
an oscillating magnetic field gradient may excite the center-of-mass oscillation of the condensate in the perpendicular direction, which will be detected more easily. Such a coupled dynamics of the condensate and non-condensate remains for a future study.

\begin{figure}
\includegraphics[width=80mm]{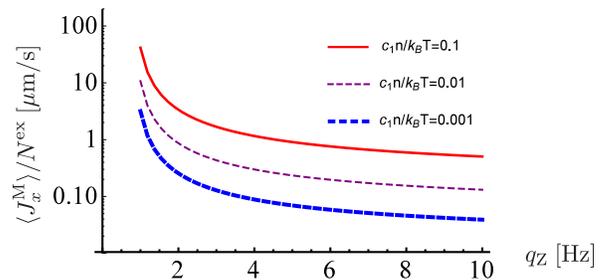}
\caption{(color online). The transverse velocity $\langle J_{x}^{M}\rangle/N^{\rm ex}$ of the excited atoms in response to the magnetic field gradient of $B^{'}$=$1.0\times 10^{-2}$mG/mm. Calculated for a spin-1 $^{23}$Na BEC with the number density $n=2.3\times 10^{20}~{\rm m}^{-3}$. Here, $N^{\rm ex}$ is the number of atoms excited in the $m=\pm1$ states, and the red (solid), purple (dashed), and blue (dash-dotted) curves show the results for $c_1n/(k_{\rm B}T)=0.1$, 0.01, and 0.001, respectively.}
\end{figure}
\section{Conclusion}
In this paper, we have shown that the MDDI, a long-range and anisotropic two-body interaction, induces the spin Hall effect in a spinor BEC.
This spin-orbit coupled transport phenomenon is mediated by thermally and quantum-mechanically excited atoms from the condensate.

We first considered the polar state of a spin-1 BEC (a condensate in the magnetic sublevel $m=0$) and investigated the properties of the excitations. We found that the spin and momentum degrees of freedom are locked in the Bogoliubov excitations. We then derived the correction of the mass- and spin-current operators due to the MDDI, where the current operators were defined from the equations of continuity. The time derivative of the mass-current operator indicates that the excited atoms in the $m=\pm1$ states feel spin-dependent Lorentz-like force. This is the origin of the spin Hall effect.

According to the linear response theory, the spin Hall conductivity $\sigma^{\rm SH}_{xy}$ is calculated from the off-diagonal correlation function between spin current and mass current. We investigated the dependence of $\sigma_{xy}^{\rm SH}$ on the strength of the MDDI, temperature, and the quadratic Zeeman energy. As expected, the conductivity becomes larger for stronger MDDI. The conductivity also becomes larger for higher temperatures. This is because the spin-orbit coupled transport is mediated by excited atoms. We also found that the conductivity remains finite even at absolute zero because of quantum fluctuations. The dependence on the quadratic Zeeman energy $q_{\rm Z}$ is more significant: The spin Hall conductivity diverges at $q_{\rm Z}=0$. This is because the magnon excitations become gapless at $q_{\rm Z}=0$ and hence an infinitesimal perturbation can excite infinite number of gapless magnons.

In a realistic situation of a spin-1 $^{23}$Na BEC, in response to a magnetic field gradient, the transverse velocity of excited atoms becomes up to 6.1~$\mu$m/s, which may be observed as an accumulation of excited atoms in the one side of the condensate. When we apply the above results to a spin-3 $^{52}$Cr BEC, the larger spin Hall conductivity is expected, where the effective strength of the MDDI compared with the spin-exchange interaction is enhanced by a factor of 5.7. For the quantitative estimation, more detailed calculation,
including the effect of the trapping potential, the finite size effect, and the coupled dynamics of the condensate and non-condensate, is required, which will be a future work.

\begin{acknowledgments}
We gratefully acknowledge Naoto Nagaosa, Ryuji Takahashi, Hiroki Isobe, Ryohei Wakatsuki, and Yusuke Sugita
for helpful discussion.
This work was supported by JSPS KAKENHI Grant Numbers 22740265 and 15K17726.
\end{acknowledgments}
\appendix
\section{THE CALCULATION OF THE OFF-DIAGONAL CURRENT-CURRENT RESPONSE FUNCTION (\ref{eq:response})}
We show how to calculate correlation function (\ref{eq:response}) in detail. We first rewrite $Q(i\omega)$ using the creation and annihilation operators of the Bogoliubov particles as
\begin{eqnarray}
&Q&(i\omega) \nonumber \\
&=&\sum_{\bf k,k'}\int_{0}^{\beta} \langle T_{\tau}\ {\hat \phi^{\dagger}_{\alpha}}(\tau){\hat \phi_{\beta}}(\tau){\hat \phi^{\dagger}_{\alpha'}}{\hat \phi_{\beta'}}\rangle e^{i\omega_{n}\tau}d\tau \nonumber \\
&\times&(\mathbf T_{\mathbf k}^{\dagger}\mathbf J_{\mathbf k,x}^{\rm M} \mathbf T_{\mathbf k})_{\alpha\beta} (\mathbf T^{\dagger}_{\mathbf k'}\mathbf J_{\mathbf k',y}^{\rm S} \mathbf T_{\mathbf k'})_{\alpha'\beta'}, \nonumber
\end{eqnarray}
where the subscripts $\alpha$ and $\beta$ take ($\pm{\bf k}, \uparrow$ or $\downarrow$),
denoting the indices for the combined  Nambu and spin space, and the repeated indices implicitly summed over. There are 6 types of expectation value $\int_{0}^{\beta} \langle T_{\tau}\ {\hat \phi^{\dagger}_{\alpha}}(\tau){\hat \phi_{\beta}}(\tau){\hat \phi^{\dagger}_{\alpha'}}{\hat \phi_{\beta'}}\rangle e^{i\omega_{n}\tau}d\tau$.
The first 4 types, 
\begin{eqnarray}
\int_{0}^{\beta}\langle T_{\tau}\ &\hat b^{\dagger}_{\alpha}(\tau)\hat b_{\beta}(\tau)&\hat b_{\alpha'}^{\dagger} \hat b_{\beta'}\rangle e^{i\omega_{n}\tau}d\tau \nonumber \\
&=&-\frac{n_{\alpha}-n_{\beta}}{i\omega_{n}+(E_{\alpha}-E_{\beta})/\hbar}\delta_{\alpha,\beta'}\delta_{\beta,\alpha'}, \nonumber \\
\label{eq:1} \\
\int_{0}^{\beta}\langle T_{\tau}\ &\hat b^{\dagger}_{\alpha}(\tau)\hat b_{\beta}(\tau)& \hat b_{\alpha'} \hat b_{\beta'}^{\dagger}\rangle e^{i\omega_{n}\tau}d\tau \nonumber \\
&=&-\frac{n_{\alpha}-n_{\beta}}{i\omega_{n}+(E_{\alpha}-E_{\beta})/\hbar}\delta_{\alpha,\alpha'}\delta_{\beta,\beta'}, \nonumber \\
\label{eq:2}\\
\int_{0}^{\beta}\langle T_{\tau}\ &\hat b_{\alpha}(\tau)\hat b_{\beta}^{\dagger}(\tau)& \hat b_{\alpha'}^{\dagger} \hat b_{\beta'}\rangle e^{i\omega_{n}\tau}d\tau \nonumber \\
&=&\frac{n_{\alpha}-n_{\beta}}{i\omega_{n}+(E_{\beta}-E_{\alpha})/\hbar}\delta_{\alpha,\alpha'}\delta_{\beta,\beta'}, \nonumber \\
\label{eq:3} \\
\int_{0}^{\beta}\langle T_{\tau}\ &\hat b_{\alpha}(\tau)\hat b_{\beta}^{\dagger}(\tau)& \hat b_{\alpha'} \hat b_{\beta'}^{\dagger}\rangle e^{i\omega_{n}\tau}d\tau \nonumber \\
&=&\frac{n_{\alpha}-n_{\beta}}{i\omega_{n}+(E_{\beta}-E_{\alpha})/\hbar}\delta_{\alpha,\beta'}\delta_{\beta,\alpha'}, \nonumber \\
\label{eq:4}
\end{eqnarray}
are the contributions from the scattering process of thermal excitations and the others,
\begin{eqnarray}
\int_{0}^{\beta}\langle &T_{\tau}& \hat b_{\alpha}^{\dagger}(\tau)\hat b_{\beta}^{\dagger}(\tau) \hat b_{\alpha'} \hat b_{\beta'}\rangle e^{i\omega_{n}\tau}d\tau \nonumber \\
&=& \frac{1+n_{\alpha}+n_{\beta}}{i\omega_{n}+(E_{\beta}+E_{\alpha})/\hbar}(\delta_{\alpha,\alpha'}\delta_{\beta,\beta'}+\delta_{\alpha,\beta'}\delta_{\beta,\alpha'}),\nonumber  \\
\label{eq:5}\\
\int_{0}^{\beta}\langle &T_{\tau}& \hat b_{\alpha}(\tau)\hat b_{\beta}(\tau) \hat b_{\alpha'}^{\dagger} \hat b_{\beta'}^{\dagger}\rangle e^{i\omega_{n}\tau}d\tau \nonumber \\
&=& -\frac{1+n_{\alpha}+n_{\beta}}{i\omega_{n}-(E_{\beta}+E_{\alpha})/\hbar}(\delta_{\alpha,\alpha'}\delta_{\beta,\beta'}+\delta_{\alpha,\beta'}\delta_{\beta,\alpha'})\nonumber, \\ \label{eq:6}
\end{eqnarray}
are from pair creation/annihilation process. We now calculate the matrix elements of $\mathbf{V}_{\mathbf k,\mu}^{\rm X} \equiv \mathbf T_{\mathbf k}^{\dagger}\mathbf J_{\mathbf k,\mu}^{\rm X} \mathbf T_{\mathbf k}$ (X$=$M or S). They can be separated into 2$\times$2 matrices as
\begin{eqnarray}
\mathbf{V}_{\mathbf{k},\mu}^{\rm X}&=&
\left(
\begin{array}{cc}
V^{\rm X}_{1,\mathbf k,\mu}& V^{\rm X}_{2,\mathbf k,\mu} \\
V^{\rm X}_{3,\mathbf k,\mu}& V^{\rm X}_{4,\mathbf k,\mu}
\end{array}
\right).
\end{eqnarray}
The elements of each matrix satisfy the following relations:
\begin{eqnarray}
(&\mathbf{V}_{1,\mathbf{k},\mu}^{\rm M}&)_{\uparrow\downarrow}=-(\mathbf{V}_{1,\mathbf{k},\mu}^{\rm M})_{\downarrow\uparrow}=(\mathbf{V}_{4,\mathbf{k},\mu}^{\rm M})_{\uparrow\downarrow}=-(\mathbf{V}_{4,\mathbf{k},\mu}^{\rm M})_{\downarrow\uparrow} \nonumber
\end{eqnarray}
\begin{eqnarray}
=\frac{1}{\hbar}\left(\frac{\partial d_{+,\mathbf k}}{\partial k_{\mu}}e^{-2i\phi}-\frac{\partial d_{-,\mathbf k}}{\partial k_{\mu}}e^{2i\phi} \right)
(u_{\mathbf k,\uparrow}u_{\mathbf k,\downarrow}-v_{\mathbf k,\uparrow}v_{\mathbf k,\downarrow}),\nonumber \\
\end{eqnarray}
\begin{eqnarray}
(&\mathbf{V}_{2,\mathbf{k},\mu}^{\rm M}&)_{\uparrow\downarrow}=-(\mathbf{V}_{2,\mathbf{k},\mu}^{\rm M})_{\downarrow\uparrow}=(\mathbf{V}_{3,\mathbf{k},\mu}^{\rm M})_{\uparrow\downarrow}=-(\mathbf{V}_{3,\mathbf{k},\mu}^{\rm M})_{\downarrow\uparrow}\nonumber
\end{eqnarray}
\begin{eqnarray}
=\frac{1}{\hbar}\left(\frac{\partial d_{+,\mathbf k}}{\partial k_{\mu}}e^{-2i\phi}-\frac{\partial d_{-,\mathbf k}}{\partial k_{\mu}}e^{2i\phi} \right)
(u_{\mathbf k,\downarrow}v_{\mathbf k,\uparrow}-u_{\mathbf k,\uparrow}v_{\mathbf k,\downarrow}),\nonumber \\ \label{eq:12}
\end{eqnarray}
\begin{eqnarray}
(\mathbf{V}_{1,\mathbf{k},\mu}^{\rm S})_{\uparrow\downarrow}=(\mathbf{V}_{1,\mathbf{k},\mu}^{\rm S})_{\downarrow\uparrow}=-(\mathbf{V}_{4,\mathbf{k},\mu}^{\rm S})_{\uparrow\downarrow}=-(\mathbf{V}_{4,\mathbf{k},\mu}^{\rm S})_{\downarrow\uparrow} \nonumber
\end{eqnarray}
\begin{eqnarray}
=\frac{2}{\hbar}\biggl\{\frac{\partial d_{5,\mathbf k}}{\partial k_{\mu}}(u_{\mathbf k,\uparrow}&-&v_{\mathbf k,\uparrow})(u_{\mathbf k,\downarrow}-v_{\mathbf k,\downarrow})\nonumber \\
&+&\frac{\partial \epsilon_{\mathbf k}}{\partial k_{\mu}}(u_{\mathbf k,\uparrow}u_{\mathbf k,\downarrow}+v_{\mathbf k,\uparrow}v_{\mathbf k,\downarrow})\biggr\}, \label{eq:13} \nonumber \\
\end{eqnarray}
\begin{eqnarray}
(\mathbf{V}_{2,\mathbf{k},\mu}^{\rm S})_{\uparrow\downarrow}=-(\mathbf{V}_{2,\mathbf{k},\mu}^{\rm S})_{\downarrow\uparrow}=-(\mathbf{V}_{3,\mathbf{k},\mu}^{\rm S})_{\uparrow\downarrow}=(\mathbf{V}_{3,\mathbf{k},\mu}^{\rm S})_{\downarrow\uparrow}\nonumber
\end{eqnarray}
\begin{eqnarray}
=\frac{2}{\hbar} \biggl\{ \frac{\partial d_{5,\mathbf k}}{\partial k_{\mu}}(u_{\mathbf k,\uparrow}&-&v_{\mathbf k,\uparrow})(u_{\mathbf k,\downarrow}-v_{\mathbf k,\downarrow}) \nonumber \\
&-&\frac{\partial \epsilon_{\mathbf k}}{\partial k_{\mu}}(u_{\mathbf k,\downarrow}v_{\mathbf k,\uparrow}+u_{\mathbf k,\uparrow}v_{\mathbf k,\downarrow})\biggr\}. \label{eq:14} \nonumber \\
\end{eqnarray}
The other components are zero.
In the definition of current operators, the matrix elements $V_{1,{\bf k},\mu}^{\rm X}, V_{2,{\bf k},\mu}^{\rm X}, V_{3,{\bf k},\mu}^{\rm X}$, and $V_{4,{\bf k},\mu}^{\rm X}$ are the coefficients of $\hat{b}^\dagger_{{\bf k},l} \hat{b}_{{\bf k},m}$, $\hat{b}_{-{\bf k},l}\hat{b}_{{\bf k},m}$, $\hat{b}^\dagger_{{\bf k},l} \hat{b}^\dagger_{-{\bf k},m}$, and $\hat{b}_{-{\bf k},l} \hat{b}^\dagger_{-{\bf k},m}$, respectively.
Hence, the contribution from the (A1)-type terms, for example, is calculated as
\begin{eqnarray}
\sum_{\mathbf k, \mathbf k'}\int_{0}^{\beta}\langle &T_{\tau}&\ \hat b^{\dagger}_{\mathbf k, m}(\tau)\hat b_{\mathbf k, l}(\tau)\hat b_{\mathbf k', m'}^{\dagger} \hat b_{\mathbf k', l'}\rangle e^{i\omega_{n}\tau}d\tau  \nonumber \\
&\times& (\mathbf{V}_{1,\mathbf{k},x}^{\rm M})_{ml} (\mathbf{V}_{1,\mathbf{k},y}^{\rm S})_{m'l'} \nonumber
\end{eqnarray}
\begin{eqnarray}
=-\sum_{\mathbf k,\mathbf k'}& &\frac{n_{\mathbf k, m}-n_{\mathbf k, l}}{i\omega_{n}+(E_{\mathbf k, m}-E_{\mathbf k, l})/\hbar}\delta_{\mathbf k,\mathbf k'}\delta_{m,l'}\delta_{l,m'}\nonumber \\
&\times& (\mathbf{V}_{1,\mathbf{k},x}^{\rm M})_{ml} (\mathbf{V}_{1,\mathbf{k},y}^{\rm S})_{m'l'} \nonumber
\end{eqnarray}
\begin{eqnarray}
=&-&\sum_{\mathbf k,\mathbf k'}(n_{\mathbf k, \uparrow}-n_{\mathbf k, \downarrow})\nonumber \\
&\times& \left[ \frac{1}{i\omega_{n}+(E_{\mathbf k, \uparrow}-E_{\mathbf k, \downarrow})/\hbar}+\frac{1}{i\omega_{n}+(E_{\mathbf k, \downarrow}-E_{\mathbf k, \uparrow})/\hbar}\right] \nonumber \\
&\times& (\mathbf{V}_{1,\mathbf{k},x}^{\rm M})_{\uparrow \downarrow} (\mathbf{V}_{1,\mathbf{k},y}^{\rm S})_{\downarrow \uparrow} \nonumber
\end{eqnarray}
\begin{eqnarray}
=&~&\sum_{\mathbf k,\mathbf k'}(n_{\mathbf k, \uparrow}-n_{\mathbf k, \downarrow})\nonumber \\
&\times& \left[ \frac{2i\hbar^{2}\omega_{n}}{-(i\hbar\omega_{n})^{2}+(E_{\mathbf k, \uparrow}-E_{\mathbf k, \downarrow})^{2}}\right] 
\nonumber \\
&\times&\frac{8\pi c_{\rm dd}n}{i\hbar} \frac{\sin{\theta}\sin{\phi}}{k}
(u_{\mathbf k,\uparrow}u_{\mathbf k,\downarrow}-v_{\mathbf k,\uparrow}v_{\mathbf k,\downarrow})\nonumber \\
&\times&\frac{\sin{\theta}\sin{\phi}}{\hbar}\biggl\{\frac{8\pi c_{\rm dd}}{3}\frac{\cos^{2}{\theta}}{k}(u_{\mathbf k,\uparrow}-v_{\mathbf k,\uparrow})(u_{\mathbf k,\downarrow}-v_{\mathbf k,\downarrow})\nonumber \\
&+&2\frac{\hbar^{2}k}{M}(u_{\mathbf k,\uparrow}u_{\mathbf k,\downarrow}+v_{\mathbf k,\uparrow}v_{\mathbf k,\downarrow})\biggr\}
\nonumber \\
=&-i&\sum_{\mathbf k,\mathbf k'}(n_{\mathbf k, \uparrow}-n_{\mathbf k, \downarrow})\nonumber \\
&\times& \left[ \frac{2i\hbar\omega_{n}}{-(i\hbar\omega_{n})^{2}+(E_{\mathbf k, \uparrow}-E_{\mathbf k, \downarrow})^{2}}\right] 
\nonumber \\
&\times&\frac{8\pi c_{\rm dd}n}{\hbar}
(u_{\mathbf k,\uparrow}u_{\mathbf k,\downarrow}-v_{\mathbf k,\uparrow}v_{\mathbf k,\downarrow})
{\sin^{2}{\theta}\sin^{2}{\phi}}\nonumber \\
&\times&\biggl\{\frac{8\pi c_{\rm dd}}{3}\frac{\cos^{2}{\theta}}{k^{2}}(u_{\mathbf k,\uparrow}-v_{\mathbf k,\uparrow})(u_{\mathbf k,\downarrow}-v_{\mathbf k,\downarrow})\nonumber \\
&+&2\frac{\hbar^{2}}{M}(u_{\mathbf k,\uparrow}u_{\mathbf k,\downarrow}+v_{\mathbf k,\uparrow}v_{\mathbf k,\downarrow})\biggr\} \nonumber \\
=&~&\sum_{\mathbf k}(n_{\mathbf k,\uparrow}-n_{\mathbf k,\downarrow}) \frac{2i \hbar \omega_{n}S(\mathbf k)}{-(i \hbar \omega_{n})^{2}+(E_{\mathbf k,\uparrow}-E_{\mathbf k,\downarrow})^{2}}\label{eq:derive1}.
\end{eqnarray}
The (A4)-type contribution has the same value as (\ref{eq:derive1}), whereas the contributions from the (A2)- and (A3)-type terms cancel with each other. 
So total contribution from scattering process of thermal excitation is the twice of (\ref{eq:derive1}), which corresponds to the first term of Eq. (34).
The contributions from (A5)- and (A6)-type terms are also calculated in the same manner:
\begin{eqnarray}
\sum_{\mathbf k, \mathbf k'}\int_{0}^{\beta}\langle &T_{\tau}&\ \hat b^{\dagger}_{\mathbf k, m}(\tau)\hat b^{\dagger}_{-\mathbf k, l}(\tau)\hat b_{-\mathbf k', m'} \hat b_{\mathbf k', l'}\rangle e^{i\omega_{n}\tau}d\tau  \nonumber \\
&\times& (\mathbf{V}_{2,\mathbf{k},x}^{\rm M})_{ml} (\mathbf{V}_{3,\mathbf{k}',y}^{\rm S})_{m'l'} \nonumber
\end{eqnarray}
\begin{eqnarray}
&=&\sum_{\mathbf k,\mathbf k'}\frac{1+n_{\mathbf k, m}+n_{\mathbf k, l}}{i\omega_{n}+(E_{\mathbf k, m}+E_{\mathbf k, l})/\hbar}\nonumber \\
&~&\times (\delta_{\mathbf k,-\mathbf k'}\delta_{m,m'}\delta_{l,l'} + \delta_{\mathbf k,\mathbf k'}\delta_{m,l'}\delta_{l,m'})\nonumber \\
&~&\times (\mathbf{V}_{2,\mathbf{k},x}^{\rm M})_{ml} (\mathbf{V}_{3,\mathbf{k}',y}^{\rm S})_{m'l'} \nonumber \\
&=& -4 \sum_{\mathbf k,\mathbf k'} \frac{1+n_{\mathbf k, \uparrow}+n_{\mathbf k, \downarrow}}{i\omega_{n}+(E_{\mathbf k, \uparrow}+E_{\mathbf k, \downarrow})/\hbar}  \nonumber \\
&~&\times (\mathbf{V}_{2,\mathbf{k},x}^{\rm M})_{\uparrow\downarrow} (\mathbf{V}_{3,\mathbf{k}',y}^{\rm S})_{\uparrow\downarrow}\nonumber \\
&=& 4i \sum_{\mathbf k,\mathbf k'} \frac{1+n_{\mathbf k, \uparrow}+n_{\mathbf k, \downarrow}}{i\omega_{n}+(E_{\mathbf k, \uparrow}+E_{\mathbf k, \downarrow})/\hbar}
\nonumber \\
&~&\times\frac{8\pi c_{\rm dd}n}{\hbar}
(u_{\mathbf k,\uparrow}u_{\mathbf k,\downarrow}-v_{\mathbf k,\uparrow}v_{\mathbf k,\downarrow})
{\sin^{2}{\theta}\sin^{2}{\phi}}\nonumber \\
&~&\times \biggl\{-\frac{8\pi c_{\rm dd}}{3}\frac{\cos^{2}{\theta}}{k^{2}}(u_{\mathbf k,\uparrow}-v_{\mathbf k,\uparrow})(u_{\mathbf k,\downarrow}-v_{\mathbf k,\downarrow})\nonumber \\
&~&+2\frac{\hbar^{2}}{M}(u_{\mathbf k,\uparrow}v_{\mathbf k,\downarrow}+v_{\mathbf k,\uparrow}u_{\mathbf k,\downarrow})\biggr\} \nonumber \\
&=& -\sum_{\mathbf k,\mathbf k'} \frac{1+n_{\mathbf k, \uparrow}+n_{\mathbf k, \downarrow}}{i\omega_{n}+(E_{\mathbf k, \uparrow}+E_{\mathbf k, \downarrow})/\hbar} 4P(\bf{k}).\label{eq:A5}
\end{eqnarray}
\begin{eqnarray}
\sum_{\mathbf k, \mathbf k'}\int_{0}^{\beta}\langle &T_{\tau}&\ \hat b_{-\mathbf k, m}(\tau)\hat b_{\mathbf k, l}(\tau)\hat b^{\dagger}_{\mathbf k', m'} \hat b^{\dagger}_{-\mathbf k', l'}\rangle e^{i\omega_{n}\tau}d\tau  \nonumber \\
&\times& (\mathbf{V}_{3,\mathbf{k},x}^{\rm M})_{ml} (\mathbf{V}_{2,\mathbf{k}',y}^{\rm S})_{m'l'} \nonumber
\end{eqnarray}
\begin{eqnarray}
&=& 4 \sum_{\mathbf k,\mathbf k'} \frac{1+n_{\mathbf k, \uparrow}+n_{\mathbf k, \downarrow}}{i\omega_{n}-(E_{\mathbf k, \uparrow}+E_{\mathbf k, \downarrow})/\hbar} (\mathbf{V}_{3,\mathbf{k},x}^{\rm M})_{\uparrow\downarrow} (\mathbf{V}_{2,\mathbf{k}',y}^{\rm S})_{\uparrow\downarrow}\nonumber \\
&=& -\sum_{\mathbf k,\mathbf k'} \frac{1+n_{\mathbf k, \uparrow}+n_{\mathbf k, \downarrow}}{i\omega_{n}-(E_{\mathbf k, \uparrow}+E_{\mathbf k, \downarrow})/\hbar} 4P(\bf{k}). \label{eq:A6}
\end{eqnarray}
The summation of these contributions reduces to the second term of Eq. (34).

\bibliography{ref}
\end{document}